\documentclass[
    ,final            
    ,numberedheadings 
  ]
  {aipproc}

\layoutstyle{6x9}

\begin{document}

\title{Gifts from Exoplanetary Transits}

\classification{97.82.-j}
\keywords      {stars: planetary systems: --- stars: rotation ---
techniques: photometric --- techniques: radial velocities ---
techniques: spectroscopic}

\author{Norio Narita}{
  address={National Astronomical Observatory of Japan, 2-21-1 Osawa,
Mitaka, Tokyo, 181-8588, Japan}
}

\begin{abstract}
The discovery of transiting extrasolar planets has enabled us
a number of interesting stduies. Transit photometry reveals
the radius and the orbital inclination of transiting planets,
and thereby we can learn the true mass and the density of
respective planets by the combined information
of radial velocity measurements.
In addition, follow-up observations of transiting planets
such as secondary eclipse, transit timing variations,
transmission spectroscopy, and the Rossiter-McLaughlin effect provide us
information of their dayside temperature, unseen bodies in systems,
planetary atmospheres, and obliquity of planetary orbits.
Such observational information, which will provide us a greater
understanding of extrasolar planets, is available only for
transiting planets.
Here I briefly summarize what we can learn from transiting planets
and introduce previous studies.
\end{abstract}

\maketitle


\section{Transit Photometry}

\subsection{Planetary Parameters}

Planetary transits in an exoplanetary system were first discovered
in 1999 for the HD~209458 system
\cite{2000ApJ...529L..45C}\cite{2000ApJ...529L..41H}.
The discovery of planetary transits enabled us to measure
the time between the start and end of transit,
the shape of ingress and egress event, and the depth of occultation.
This allows us to estimate several planetary parameters
for transiting systems by modeling transit light curves.
There are some (almost) equivalent but different types of formalism
describing transit light curves
\cite{2002ApJ...580L.171M}
\cite{2006A&A...450.1231G}
\cite{2009ApJ...690....1O}.
Based on these formulae, transit light curves are characterized by
the combination of following parameters:
the ratio of star-planet radii $R_p/R_s$,
the semimajor axis in units of the stellar radius $a/R_s$,
the orbital inclination $i$,
the limb-darkening parameters (e.g., $u_1$ and $u_2$ for the
quadratic limb-darkening law),
the time of mid-transit $T_c$,
the phase of igress or egress $\theta$, and
the transit duration $\tau$.
By combining with the information from radial velocity observations,
one can determine the true planetary mass $M_p$ and also the density
of the planet $\rho$ for transiting planets.
In addition, the mass-radius distribution enables us to estimate
the planetary interior structure by modeling planetary atmospheres.

\subsection{Transit Surveys and Follow-up Observations}

After the first discovery of transit events, several teams started
programs of transit surveys with ground-based telescopes
\cite{2002AcA....52....1U}
\cite{2004ApJ...613L.153A}
\cite{2006ApJ...648.1228M}
\cite{2007MNRAS.375..951C}
\cite{2007ApJ...656..552B}.
Thanks to these ground-based transit surveys, the number of detections
of transiting planets dramatically increased in these years.
On the other hand, it is also important to conduct precise follow-up
observations to determine precise parameters of transiting planets.
For example, such follow-up observations are imperative to confirm
the existence fo "inflated" planets (e.g., HD209458b) or planets
with massive cores (e.g., HD149026b).
Some groups have focused on such follow-up studies
(e.g., the Transit Light Curve project
\cite{2006ApJ...652.1715H}\cite{2007AJ....133...11W}).
and have contributed to improve the knowledge of exoplanetary nature.
Such ground-based transit surveys and follow-up observations are
nowadays actively conducted in the world, and also two spectacular
space missions \textit{CoRoT} \cite{2006cosp...36.3749B} and
\textit{Kepler} \cite{1997ASPC..119..153B} have been successfully initiated.
These space-based transit surveys would open the door for detections
of smaller Earth-like transiting planets in the near future.

\subsection{Thermal Emission}

In infrared region, the planet-to-star flux ratio is higher compared to
that in optical wavelength. The thermal emission from a hot planet is
detectable when the planet passes behind the host star, namely
the time of secondary eclipses.
The first detections of secondary eclipses were reported for
HD209458b and TrES-1b with the \textit{Spitzer Space Telescope} in 2005
\cite{2005ApJ...626..523C}
\cite{2005Natur.434..740D}.
The detections of secondary eclipses provide us the information
of planetary dayside temperature.
Subsequently, Knutson et al. (2007) \cite{2007Natur.447..183K}
presented a series of \textit{Spitzer} observations of HD189733b
over a half of the planetary orbital period, including a planetary
transit and a secondary eclipse.
They detected phase variations of HD189733b and reported a temperature
map (day-night contrast) of the exoplanet.
In addition, multi-wavelength secondary eclipse observations
for several transiting hot Jupiters revealed that some of hot Jupiters
have temperature inversion in upper atmosphere
\cite{2008ApJ...673..526K}
\cite{2009ApJ...691..866K}.
Also, very recently a few groups reported ground-based detections of
secondary eclipses
\cite{2009A&A...493L..31S}
\cite{2009A&A...493L..35D}.
In the coming decade, larger infrared space telescopes such as
\textit{JWST} and \textit{SPICA} will enable us to study thermal emission
of smaller and colder (but still warm) transiting extrasolar planets.

\subsection{Transit Timing Variations}

Given that a transiting planet is the one and only body in
the planetary system, transit events would occur at a regular interval
(except for the effect of general relativity and stellar quadrupoles).
However, if there is any additional bodies around transiting planets,
such as other planets or moons, the interval of consecutive transits
is not constant in principle.
Such variations is referred to as ``transit timing variations''
(hereafter TTVs). Theoretical studies for TTVs caused by another
unseen planet were presented
\cite{2005Sci...307.1288H}
\cite{2005MNRAS.359..567A}, and TTVs caused by an exomoon around
a transiting planet were studied
\cite{2006A&A...450..395S}
\cite{2007A&A...470..727S}
\cite{2009MNRAS.392..181K}.
An interesting point of observations of TTVs is that a typical
amplitude of TTVs is large enough to detect an Earth-mass planet
around a transiting hot Jupiter even with ground-based telescopes
if the unseen planet is in mean motion resonance.
Recently a detection of TTVs for OGLE-TR-111b was reported
\cite{2008ApJ...682L..49D}.
Observational studies of TTVs will become more active
in the CoRoT/Kepler era.

\clearpage
\section{Spectroscopic Studies of Transiting Planets}

\subsection{Transmission Spectroscopy}

\vspace{-5mm}
\begin{figure}[pth]
  \includegraphics[height=.2\textheight]{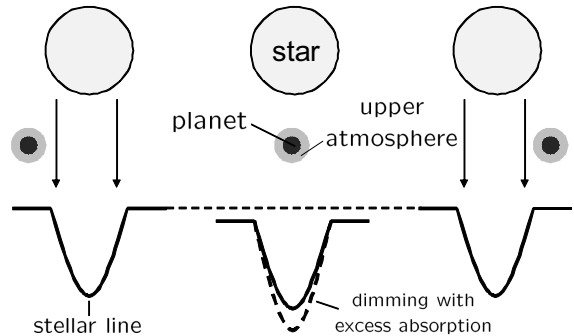}
  \caption{An illustration for the concept of transmission spectroscopy.}
\end{figure}

During a planetary transit, a small fraction of the
stellar surface is blocked by the planetary body, and
even smaller fraction of the starlight is transmitted through
the optically thin portion of the planetary upper atmosphere.
Thus for transiting planets, one can search for additional absorption
features due to the planetary atmosphere by obtaining spectra with
sufficiently high spectral resolution and high signal-to-noise ratio,
and comparing the spectra taken in and out of transit,
at least in principle.
This methodology is referred to as ``transmission spectroscopy''
(Figure~1).

Pioneering theoretical studies predicted that such excess absorption
would be especially strong in the optical transitions of alkali metals,
such as Na~{\sc i} ($\sim$5890, $\sim$5896\AA)
and K~{\sc i} ($\sim$7670\AA)
\cite{2000ApJ...537..916S}\cite{2001ApJ...553.1006B}.
In particular, the signal for the sodium doublet lines
(predicted by cloudless models) was as strong as $\sim 0.1\%$
and the strongest in optical wavelength, and thus the sodium lines
were considered to be the best target for this study.

Later, the first detection of an additional $0.0232 \pm 0.0057\%$
absorption for the sodium doublet lines ($5893 \pm 6$\AA)
was made for the HD209458 system with the STIS on board
the Hubble Space Telescope (\textit{HST}) \cite{2002ApJ...568..377C}.
However the size of the signal was considerably weaker than
the original prediction of $\sim 0.1\%$ with the cloudless models,
and the weakness of the absorption has prompted theoreticians to
invoke the presence of high cloud decks or haze in this planet.

On the ground side, however, some observers tried to detect the excess
absorption in HD209458b with ground-based large telescopes,
but cound not achieve sufficient sensitivity
\cite{2000PASP..112.1421B}
\cite{2001A&A...371..260M}
\cite{2005PASJ...57..471N}.
However, recently a few groups report detections of
sodium excess absorption in HD209458b and also HD189733b
\cite{2008ApJ...673L..87R}
\cite{2008A&A...487..357S}.
The reports would stimulate independent confirmation and further
ground-based studies in the near future.

On the other hand, transmission spectroscopy has been also conducted
in infrared wavelength region with the \textit{Spitzer} telescope,
and some observers succeeded in detecting
a few kinds of molecules in atmospheres of hot Jupiters
(e.g., \cite{2007Natur.448..169T}\cite{2008Natur.452..329S}).

The methodology of transmission spectroscopy would be especially
important in the future when larger infrared space telescopes
\textit{JWST} and \textit{SPICA} are available, in order to search for
bio-signatures in transiting Earth-like or Super-Earth planets
discovered by space-based transit survey missions such as
\textit{CoRoT}, \textit{Kepler}, and \textit{TESS},
although such transmission spectroscopy observations may be difficult
\cite{2008SPIE.7010E..51T}.

\subsection{The Rossiter-McLaughlin Effect}

\begin{figure}[pth]
  \includegraphics[height=.2\textheight]{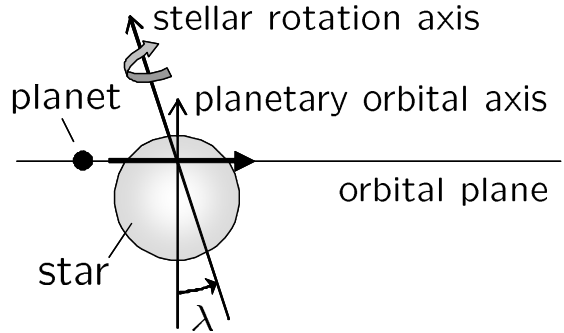}
  \caption{The sky-projected spin-orbit alignment angle $\lambda$.}
\end{figure}

The Rossiter--McLaughlin effect (hereafter the RM effect)
is a phenomenon originally reported as a
``rotational effect'' in eclipsing binary systems by
Rossiter (1924) \cite{1924ApJ....60...15R} (for the Beta Lyrae system) and
McLaughlin (1924) \cite{1924ApJ....60...22M} (for the Algol system).
In the context of transiting planets, Queloz et al. (2000)
\cite{2000A&A...359L..13Q} first reported the detection of
the RM effect in the HD209458 system.
Subsequently, theoretical discussions and analytic formulae for
the RM effect were studied
\cite{2005ApJ...622.1118O}
\cite{2006ApJ...650..408G}
\cite{2007ApJ...655..550G}.

The RM effect is seen as the radial velocity anomaly during
a transit caused by the partial occultation of the rotating
stellar surface.
Since the radial velocity anomaly depends on the trajectory of
the planet across the host star, by monitoring this
anomaly throughout a transit, one can determine the sky-projected angle
between the stellar spin axis and the planetary orbital axis denoted
by $\lambda$ (see Figure~2).

The angle $\lambda$ has become widely recognized to be an useful
diagnostic to investigate planet migration histories.
It is because the standard disk-planet interaction models (Type II
migration models) generally predict small misalignments between
the stellar spin axis and the planetary orbital axis, while
planet-planet scattering models 
\cite{2008ApJ...678..498N}
\cite{2008ApJ...686..580C}
or the Kozai migration models for planets in binary systems
\cite{2003ApJ...589..605W}
\cite{2005ApJ...627.1001T}
\cite{2007ApJ...669.1298F}
\cite{2007ApJ...670..820W}
have possibilities to produce large spin-orbit misalignments.

So far, the spin-orbit alignment angle $\lambda$ has been measured
for over 10 transiting planets
\cite{2005ApJ...631.1215W}
\cite{2006ApJ...653L..69W}
\cite{2007PASJ...59..763N}
\cite{2007ApJ...665L.167W}
\cite{2007ApJ...667..549W}
\cite{2008A&A...481..529L}
\cite{2008PASJ...60L...1N}
\cite{2008A&A...488..763H}
\cite{2008ApJ...683L..59C}
\cite{2008ApJ...682.1283W}
\cite{2008ApJ...686..649J}
\cite{2008arXiv0812.0785B}
\cite{2009arXiv0902.3461W}
\cite{Narita2009}.
As evidenced by the number of recent publications,
measurements of the RM effect have become
one of the most active research topics in exoplanetary science.

Among the observed transiting planets, three transiting eccentric planets
(HAT-P-2b, XO-3b, and HD17156b) are particularly interesting targets,
because the migration models that produce
eccentric orbits may also produce large spin-orbit misalignments.
Interestingly, the outcomes of the measurements of these systems
were different; HAT-P-2b and HD17156b are well-aligned
and the results resemble the majority of other non-eccentric systems
for which the RM effect has been measured,
while XO-3b has a significant spin-orbit misalignment and
it is currently the one and only exception which has a large
spin-orbit misalignment.

However, since we cannot draw robust conclusions with just
one sample of a spin-orbit misalignment \cite{2009arXiv0902.0737F},
further observations of the RM effect are desired to investigate
migration mechanisms of extrasolar planets in the future.

\clearpage
\begin{center}
{\bf Questions and Answers at the Conference}\\
\end{center}
{\it Question by D. N. C. Lin}:\,
Is there any possibility to detect weather variation of
transiting planets by transmission spectroscopy?
\smallskip \\
{\it Answer}:\,
Weather variations may cause variations of excess absorption level,
since haze or high cloud decks may suppress such absorption.
Thus if we are able to detect excess absorption for a single transit,
we can compare the excess absorption level for every transit
and can search for weather variations in transiting exoplanets.
\quad \medskip \\
{\it Question by E. Kokubo}:\,
How can we detect planetary rings around transiting planets?
\smallskip \\
{\it Answer}:\,
Precise photometric monitoring of planetary transits will enable us
to search for rings \cite{2004ApJ...616.1193B}. Also, precise
measurements of the Rossiter-McLaughlin effect may provide us
independent evidence of planetary rings \cite{2009ApJ...690....1O}.
\quad \medskip \\
{\it Question by E. L. Turner}:\,
Does the XO-3b result confirm migration scenarios of planet-planet
scattering or the Kozai models statistically?
\smallskip \\
{\it Answer}:\,
The existence of spin-orbit misaligned planetary systems favors
dynamical scenarios of planet-planet scattering or
the Kozai migration models, and so the XO-3b result is an important
result to confirm such scenarios \cite{2009arXiv0902.0737F}.
However, since currently XO-3b is the one and only system that have
a large spin-orbit mislignment, more observations of
the RM effect are desired, especially for transiting eccentric planets
or transiting planets in binary systems, so as to detect more
spin-orbit misaligned planets.


\begin{theacknowledgments}
I acknowlege the support by a Japan Society for Promotion of Science
(JSPS) Fellowship for Research (PD: 20-8141) and 
Grants-in-Aid for Scientific Research on Priority Areas,
``Development of Extra-Solar Planetary Science,'' from the Ministry
of Education, Calture, Sports, Science, and Technology
(16077101, 16077204).
\end{theacknowledgments}





\end{document}